\documentclass[fleqn,usenatbib]{mnras}

\usepackage{newtxtext,newtxmath}

\usepackage[T1]{fontenc}

\DeclareRobustCommand{\VAN}[3]{#2}
\let\VANthebibliography\thebibliography
\def\thebibliography{\DeclareRobustCommand{\VAN}[3]{##3}\VANthebibliography}

\usepackage{graphicx}	
\usepackage{amsmath}	

\renewcommand{\[}{\begin{equation}}
\renewcommand{\]}{\end{equation}}
\def\p{\partial}\def\i{{\rm i}}

\let\boldgrk=\gkvecten
\let\boldgrksc=\gkvecseven

\def\gkthing#1{{\mathchoice%
	{\hbox{{\boldgrk\char#1}}}
	{\hbox{{\boldgrk\char#1}}}
	{\hbox{{\boldgrksc\char#1}}}
	{\hbox{{\boldgrksc\char#1}}}}}

\def\vtheta{\gkthing{18}}

{\newif\ifnotend
\notendtrue
\def\veclist{ABCDEFGHIJKLMNOPQRSTUVWXYZabcdefghijklmnopqrstuvwxyz.}
\def\top#1#2.{#1}
\def\tail#1#2.{#2.}
\loop\expandafter\xdef\csname v\expandafter\top\veclist\endcsname%
{{\noexpand\bf\expandafter\top\veclist}}
\edef\veclist{\expandafter\tail\veclist}
\if\veclist.\notendfalse\fi\ifnotend\repeat}
{\newif\ifnotend
\notendtrue
\def\callist{ABCDEFGHIJKLMNOPQRSTUVWXYZ.}
\def\top#1#2.{#1}
\def\tail#1#2.{#2.}
\loop\expandafter\xdef\csname c\expandafter\top\callist\endcsname%
{{\noexpand\cal\expandafter\top\callist}}
\edef\callist{\expandafter\tail\callist}
\if\callist.\notendfalse\fi\ifnotend\repeat}

\def\la{\langle}

\def\d{{\rm d}}

\def\apjs{ApJS}
\def\apjl{ApJL}

\def\aap{A\&A}

\def\vnabla{{\bf\nabla}}

\def\fracj#1#2{{\textstyle{#1\over#2}}}

\title[Ensemble Averages]{Boltzmann Equation Field Theory I: Ensemble Averages}

\author[Lau Jun Yan]{
Jun Yan Lau,$^{1,2}$\thanks{E-mail: jundoesphysics@gmail.com}
\\
$^{1}$Mullard Space Science Laboratory, University College London, Holmbury House, Holmbury Hill Road, Dorking, RH56NT UK\\
$^{2}$Tsung-Dao Lee Institute, Shanghai Jiao Tong University, 1 Lisuo Road, Pudong New Area, Shanghai 201210 China
}

\date{Accepted XXX. Received YYY; in original form ZZZ}

\pubyear{\the\year{}}

\begin{document}
\label{firstpage}
\pagerange{\pageref{firstpage}--\pageref{lastpage}}
\maketitle

\begin{abstract}
I present an unbiased method of mapping particles to distribution functions and vice versa. This method alone encapsulates the canonical formulation of statistical mechanics, since it can be used to derive the principle of maximum entropy in both Boltzmann's paradigm and Gibbs' paradigm. A rigorous definition of the macrostate enables application of this statistical mechanical theory to self-gravitating systems, by decoupling time-averages and ensemble averages. I compute two-point correlation functions for self-gravitating and electrostatic systems.
\end{abstract}




\section{Introduction}

The\footnote{This introduction is taken from my thesis introduction \citep{LauJunYan2024}.} goal of statistical mechanics is to connect the macroscopic features of systems with the microscopic interactions between the particles that comprise them. In the context of astrophysical dynamics, the macroscopic features are the morphologies and dynamics of globular clusters and galaxies, while the microscopic interactions are the gravitational forces between the stars that comprise these systems. 

So how do we connect the macroscopic and microscopic features of self-gravitating systems? The statistical mechanics of self-gravitating systems will be an important tool to describe the out-of-equilibrium, unsmooth nature of modern observations of astrophysical systems. 

To begin with, let us ask the question: What are microscopic and macroscopic features? 

A microstate is a list of positions and velocities, phase-space coordinates $\vw = (\vx,\vv)$ of length $N$, $\{\vw_i\} \equiv \{\vw_1,\vw_2,...,\vw_N\}$ describing a system at a single point in time.\citep[This is what][refers to as the ``Komplexion''.]{Boltzmann1877} 

A macrostate on the other hand, is defined in a self-referential manner. A macrostate is defined as a complete list of macroscopic variables---variables that are thermodynamically relevant to a system, while a theory of thermodynamics is defined by how it relates one macroscopic variable to another; as exemplified by the first law of thermodynamics \citep{Clausius1850}; $\Delta U = Q - W$ that relates the change in the internal energy $\Delta U$ of a closed system with the difference of the heat $Q$ introduced into the system and the thermodynamic work $W$ done by the system. The reason macroscopic variables are defined in this cyclical manner is that thermodynamics was developed first as a phenomenological theory, before it was described by statistical mechanics. Defining the concept of a macroscopic feature thus requires the inspection of a thermodynamical theory.

Classical thermodynamics, which is explained by the microscopic interpretation provided by Boltzmann-Gibbs statistical mechanics (BGSM), prescribes relationships between macroscopic quantities such as pressure, temperature and work done under certain assumptions. 

These macroscopic quantities all share a common definition that was first captured by \cite{Bernoulli1738} (in his Hydrodynamica, Chapter 10, Sections 4 and 6, where he outlines the kinetic theory of gases) who singled out a class of macroscopic quantities by focusing on the ones that can be understood by time-averaging their corresponding microscopic quantities---pressure from taking a time average of the momentum transferred from gas particles hitting the walls of a box, for example. A calculation like this, however, requires integrating an initial condition of microstates forwards in time, and is intractable for interacting systems.

It was \cite{Gibbs} who explicitly replaced the deterministic but chaotic production of microstates sourced from time-evolution with a stochastic alternative. He posits that on the lengthy time-scales over which Bernoulli's macroscopic quantities are measured, the particles have had sufficient opportunity to `rearrange' themselves, their time-evolution sampling all the microstates available to them with equal probability: the so-called ergodic hypothesis. 

The ergodic hypothesis is a justification and mathematical encoding of one of Boltzmann's postulates: that the dynamics of microstates functions to chaotically (in the classical sense of mixing) `scramble' information that is inherent to the microstate. This mixing conserves only the collisional invariants while maximising uncertainty regarding our knowledge of the microstates; a postulate he used to derive the Boltzmann (thus, Maxwell-Boltzmann) distribution. This postulate (and thus the ergodic hypothesis) is well suited for a gas with particles that exhibit short-ranged interactions, where interactions deflect particles from their original trajectories thus `scrambling' the system but conserving total momentum and energy. However, does it describe systems with strong, long-ranged interactions? Gibbs certainly did not think so, thus he (and Bernoulli both) provided an additional assumption: that the energy of the system $E[\{\vw_i\}] = \sum_i E(\vw_i)$ could be expressed as a sum of the energy of each particle within the microstate, or that systems described by BGSM had to be composed of weakly-interacting particles.

The ergodic hypothesis aligns with the intuition that one should only measure the pressure of a system by summing over a large number of collisions---so as to suppress noise fluctuations in the collision rate associated with the stochastic nature of particles. It not only does away with the need to tackle (weak) dynamics, but also removes the need for a microstate. Hence it is understood that BGSM only applies for systems that are at equilibrium/adiabatically changing; that is systems for which the macroscopic variables change far more slowly than the time taken for a particle to make its rounds within the system, which is a function of the thermal speed and the size of the system considered.

What fundamentally prevents applications of BGSM to gravitating systems is not just that it only applies to weakly interacting systems with short ranged forces (the gravitational force is long-ranged), or just that the ergodic hypothesis does not function (the surface of constant energy defined in the space of microstates is unbounded in self-gravitating systems, and hence the ergodic hypothesis fails) \citep[for more reasons, see][Box 7.1]{GDII}, but it is that the macroscopic features we astrophysicists are interested in are not described by BGSM.

When an inherently chaotic (but weakly interacting) system evolves for a sufficiently long period, time averages naturally equate to ensemble averages, which causes measurements that are taken over such periods to correspond to ensemble averages (i.e. a measurement of temperature). However, does this picture align with the way we observe stars? Observations of stars within our Milky Way are made within the slightest of instants, relative to astrophysical timescales. We are in radically different regimes to Gibbs and Bernoulli: our macroscopic features are not ones which are persistent such that they evolve only across secular timescales, but rather are system-scale phase-space fluctuations---They are collective motions in the microstate: spirals, bars, and dipole asymmetries that do not belong under the category of Bernoulli's macroscopic quantities. This is the difference between a measurement of the velocity-dispersion and a measurement of the temperature of a system: measuring the latter implies stationarity; while you can measure the velocity-dispersion at any point in time.

To address the question that kickstarted this investigation into thermodynamics, I propose that a macroscopic feature is a common feature found amongst all representative models of a system---this captures both temporally persistent features and features with strong phase-space signatures and removes human bias in defining macroscopic quantities. 

This novel definition of a macroscopic feature allows us to refine the definition of the macrostate. The macrostate is thus defined as a complete list of all common features found amongst all representative models of a system, and is therefore the distribution of all representative models of a system. 

Now we have one question left to answer: How do we (representatively) model an $N$-particle self-gravitating system?

Section 2 describes the mathematical methods used in the formulation of this theory. Section 3 motivates the notion of proper representation, and describes how we can representatively model $N$-particle self-gravitating systems, and Section 4 explains how to obtain macroscopic quantities from this theory. Section 5 discusses the ramifications of this theory, and Section 6 concludes.
 
\section{Mathematical Methods}

In this section I detail the mathematical methods and assumptions employed in this paper. 

\subsection{Poisson Sampling}

This paper is focused on the study of the statistical mechanics of statistically equivalent particles. By this I mean that they are independently sampled from the same number density, $f$. Further to that, I assume that the particles only differ in their positions and velocities, the phase-space coordinates $\vw = (\vx,\vv)$. This means that $f = f(\vw,t)$, where $t$ is time, and all other parameters---charge, mass, spin---are shared between particles. 

I assume that the particles are Poisson sampled from a number density $f$. A single Poisson sampling of $f$ produces particles by holding Bernoulli trials at every point in phase-space, where the probability of successfully locating a particle at $\vw$ is the infinitesimal $f(\vw) \d^6\vw$ and the probability of failing to do so is $1 - f(\vw) \d^6\vw$. 

This means that the probability of sampling a realisation of $N$ particles at the coordinates $(\vw_1,\vw_2,...,\vw_N) = \{\vw_i\}$ is,
\[
p^{(N)}(\{\vw_i\},t) = \prod_{i=1}^N \Big(f(\vw_i,t) \d^6\vw_i\Big)
\]
where we neglect the probabilities of not finding particles since they are $\approx 1$, and the $N$-particle probability density is,
\[\label{eq:NptProbDens}
f^{(N)}(\{\vw_i\},t) = \prod_i f(\vw_i,t) = \prod_i f_i.
\]
We will use the subscript notation $f_i = f(\vw_i,t)$ to denote the single-particle probability evaluated at the position $\vw_i$ for the rest of this paper. Where there is no subscript, reference is instead made to a particle at a generic phase-space coordinate $\vw$.

One key aspect of Poisson sampling is that the number of particles per sample is not fixed. To illustrate this, we compute the expected number of particles in a single sampling of $f$, $\overline{n}$, and its variance $\overline{(n - \overline{n})^2}$.
\[\begin{aligned}\label{eq:ExptN}
\overline{n} &= \int \d^6\vw~f = \mu,\\
\overline{(n - \overline{n})^2} & = \int \d^6\vw~ f (1 - f \d^6\vw) = \int \d^6\vw~f = \overline{n},
\end{aligned}\]
where \[
\mu = \int \d^6\vw~f
\] 
is the number of particles encoded by $f$ and we have used Bernoulli statistics, i.e. the mean is $p$ and the variance is $p(1-p)$ where $p$ is the probability of success. Poisson sampling applied to individual points in phase-space produces Poisson statistics across all of phase-space.

While $f$ is a number density, the way we draw from it is such that each and every phase-space position is sampled independently from every other phase-space position. It is thus possible to sample all available phase-space and obtain a number of particles that differs from $\mu$. 

This counterintuitive result can be understood as a reshuffling of the underlying positions of particles between the drawing of each sample, such that failures and successes in locating particles do not increase or decrease the probability that the particle is elsewhere, respectively. More colloquially, this is ``sampling with replacement''. This feature makes my treatment of $f$ more akin to an unnormalised probability distribution.

An alternate interpretation of this deviation from normalisation is simply that $f$ in this theory plays the role of the model to an $N$-particle system, and the model may deviate from reality, at the cost of being assigned a lower probability.

\subsection{The Law of Large Numbers}

The Law of Large Numbers describes how averages of some function $g(\vw)$ with respect to $f(\vw)$ can be taken by sampling $g(\vw_i)$ from $f$ repeatedly. 

Note this differs slightly from Monte Carlo Integration \citep{MetropolisUlam1949}, which applies not to an $N$-fold sampling, but rather a distribution of $N$ particles. For $N$ particles that are assumed to have been sampled from $f$, we find:
\[
\lim_{N \rightarrow \infty}{1 \over N}\sum_{i=1}^{N} g(\vw_i) = {1 \over \mu}\int \d^6\vw~ gf.
\]
The Law of Large Numbers applied to an $\tilde{N}$-fold sampling of $g(\vw_i)$ from $f$ produces (equation \eqref{eq:ExptN}) $\tilde{N}\mu$ particles, and the expectation of $g$ taken with respect to each sample is:
\[
\lim_{\tilde{N} \rightarrow \infty} {1 \over \tilde{N}}\sum_{i = 1}^{\tilde{N}\mu} g(\vw_i) = \int \d^6\vw~ gf.
\]

\subsection{From Liouville's to Boltzmann's}

In this subsection, I present a derivation of the collisionless Boltzmann equation (CBE) that differs assumption-wise from truncating the \cite{BornGreen1946,Bogoliubov1946,Kirkwood1946,Yvon1935} (BBGKY) hierarchy. 

\cite{Liouville}'s equation (found in \cite{GDII}) describes the time evolution of an arbitrary $N$-particle distribution function, $f^{(N)}(\vw_1,\vw_2,...,\vw_N,t) = f^{(N)}(\{\vw_i\},t)$, \[{\p f^{(N)} \over \p t} + \sum_i \Bigg[f^{(N)},H^{(N)}\Bigg]_i = 0\] within an $N$-particle Hamiltonian (we employ a self-gravitating Hamiltonian), 
\[
H^{(N)}(\{\vw_i\}) = \sum_{i=1}^N \Bigg(\fracj12 m_i\vv_i^2 - \sum_{j\neq i, j = 1}^N \fracj12 {Gm_i m_j \over |\vx_i-\vx_j|}\Bigg).
\]
where the Poisson brackets are
\[[A,B]_i = {\p A\over \p \vq_i}\cdot{\p B\over \p \vp_i}- {\p A \over \p \vp_i}\cdot{\p B \over \p \vq_i}\] and $\vw_i = (\vq_i,\vp_i) = (\vx_i,\vv_i)$ are the canonical (position, velocity) coordinates of the $i$-th particle. 

We now apply it---not to a known $N$-particle distribution function, but the distribution function of the expected outcome of a $\tilde{N}$-fold Poisson sampling of the one particle DF, $f$. The expected number of particles is $\tilde{N}\mu$ (equation \eqref{eq:ExptN}), and we find
\[
\sum_{i=1}^{\tilde{N}\mu} \prod_{k\neq i} f_k \Bigg({\p f_i \over \p t} + m \vv\cdot {\p f_i\over \p \vx_i} + {\p f_i\over \p \vv_i}\cdot{\p \over \p \vx_i} \sum_{j \neq i}^{N\mu} {G  m^2 \over |\vx_i - \vx_j|}\Bigg) = 0,
\]
where $\vw = (\vx,\vv)$, the positions and velocities, are canonical coordinates. We cannot set the product $\prod_{k\neq i} f_k = 0$, because $\{\vw_i\}$ are the locations of particles, and particles cannot be sampled in regions with zero probability, thus we find
\[\label{eq:PrecursorCBE}
{\p f_i \over \p t} + m \vv_i \cdot {\p f_i \over \p \vx_i} - {\p f_i \over \p\vv_i} \cdot {\p \over \p \vx_i} \Bigg(-\sum_{j\neq i}^{\tilde{N}\mu}{G m^2 \over |\vx_i - \vx_j|}\Bigg) = 0.
\]
The sum in the large brackets can be resolved via the law of large numbers. Poisson sampling statistics imply that for large $\tilde{N}$,
\[\begin{aligned}\label{eq:LargeNoPot}
{-{1 \over \tilde{N}}\sum_{j\neq i}^{\tilde{N}\mu} {Gm^2\over |\vx_i - \vx_j|}} &\rightarrow -{\tilde{N}\mu-1 \over \tilde{N}\mu} \int \d^6\vw' {Gm^2 \over |\vx_i - \vx'|} f(\vw')\\
&\rightarrow  -\int \d^6\vw~ {Gm^2 \over |\vx_i - \vx'|} f(\vw').
\end{aligned}\]
Substituting equation \eqref{eq:LargeNoPot} into equation \eqref{eq:PrecursorCBE} gives us the Collisionless Boltzmann Equation,
\[\label{eq:CBE}
{\p f \over \p t} + [f,H] = 0,
\]
where the one-particle Hamiltonian is,
\[
H = \fracj12 m \vv^2 + m \Phi(\vx),
\]
$\tilde{N}*m = M$, the total particle mass and the potential
\[\label{eq:Phi}
\Phi[f](\vx) = -\int \d^6\vw' {GM \over |\vx - \vx'|} f(\vw').
\]
This derivation necessitates the use of the law of large numbers, and is completed by assuming that the number of samples $\tilde{N}$ is equal to the number of particles in the observed system for which $f^{(N)}$ is defined, $N$. Note that Poisson sampling statistics means that equating the number of samples to the number of particles does not imply that $\mu = 1$, thus this derivation extends the CBE to unnormalised number densities.

What separates the CBE from Liouville's equation is how the particles move in the potential of the distribution function, $\Phi = \Phi[f]$, that is the expected potential, not the true potential that is described by the phase-space positions of each particle. Making a choice of $f$ that reflects the positions of particles $\{\vw_i\}$ is thus essential to the consistency of the CBE with Liouville's equation. 

\subsection{Self-consistent Hamiltonian}

The Hamiltonian, $H$ is the energy of a star with mass $m$ travelling in a gravitational potential $\Phi$ is
\[
H = \fracj12 m \vv^2 + m\Phi(\vx,t).
\]
When this star moves in the potential created by a mass density $M*f(\vw,t)$, the potential is a functional of $f$ that takes the form in equation \eqref{eq:Phi}.

Astrophysical systems of interest are self-gravitating; that is they evolve under their own gravitational forces. Such a system with mass density $M*f$ conserves the self-consistent energy,\[\label{eq:SelfConsistentE}
E[f](t) = \int \d^6\vw~ f\Bigg(\fracj12 m \vv^2  + \fracj12 m  \Phi[f]\Bigg)
\]under the action of the CBE. 

Observe,\[\begin{aligned}\label{eq:GlobaltoLocal}
{\p \over \p t}E[f](t) &= \int \d^6\vw~ {\delta E(t) \over \delta f(\vw,t)} \Bigg({\p f(\vw,t) \over \p t }\Bigg)\\
&= \int \d^6\vw~ H\Bigg(-[f,H]\Bigg)\\
&= \int \d^6\vw~f[H,H] = 0.
\end{aligned}\] where I have used integration by parts between the second and final equalities and assumed, as it is customary, that $f$ diminishes to zero at the boundaries of phase-space.

\subsection{Functional Differentiation}
Central to the calculations in this paper is the use of functional analysis. 

Functional differentiation can be understood as a part of the calculus of variations. The variation in $E[f]$ by varying $f$ by a small $\delta f$ is to leading order in $\delta f$,
\[
\delta E[f,\delta f] = \lim_{\epsilon \rightarrow 0}{E[f + \epsilon\delta f] - E[f] \over \epsilon} = \int \d^6\vw~ {\delta E \over \delta f(\vw)}\delta f(\vw).
\]
Using this definition of the functional derivative, we find that the definition
\[
{\delta E[f, \delta f] \over \delta f(\vw')} \equiv {\delta E \over \delta f(\vw)} 
\]
implies
\[\label{eq:FunctDerivDef}
{\delta f(\vw,t) \over \delta f (\vw',t)} = \delta^6(\vw - \vw')
\]
meaning a simpler way to understand functional derivatives is to consider how how a functional would be perturbed by the introduction of a Dirac delta function,
\[
{\delta E[f(\vw)] \over \delta f(\vw')} = \lim_{\epsilon \rightarrow 0} {E[f + \epsilon \delta^6(\vw - \vw')] - E[f]\over \epsilon}.
\]
Identifying $E[f]$ with the self-consistent energy (equation \eqref{eq:SelfConsistentE}), we find
\[\begin{aligned}\label{eq:EvarH}
{\delta E \over \delta f(\vw')} &= \int \d^6\vw~ \fracj12 m \vv^2 \delta^6(\vw - \vw') \\&+ \fracj12 m \delta^6(\vw - \vw') \Phi[f](\vx) \\&+ m f \int \d^6\vw_a \Bigg(-{GM \over |\vx - \vx'|} \delta^6(\vw_a - \vw')\Bigg)\\
&=\fracj12 m{\vv'}^2 + m\Phi[f](\vx',t) = H(\vw',t).
\end{aligned}\]
The functional derivative of $E$ thus describes how the change in the total energy of a system in response to the addition of a single particle at a point in phase-space must be equal to the single particle energy at that point; that is strictly the Hamiltonian in the context of gravitating systems. 

\subsection{Functional Integration}

The functional integral is the inverse of the functional derivative. 
\[
\int \mathcal{D}f = \int_{\mathbb{R}}... \int_{\mathbb{R}} \prod_{\vw} \d f(\vw)
\]
Functional integration over $f$ is understood as integrating over all possible values of $f$ at every point in phase-space $\vw$ where $f$ is defined, that is the integration over all possible $f$s at a fixed point in time. 

The ability to functionally integrate over $f$ is essential to trying to understand how we can connect particles with the distributions they are sampled from, given that the map between distributions and particles is many-to-many. 

\section{Assigning distributions to particles}

How can we assign a distribution function to a single sample of $N$ particles, $\{\vw_i\}$?

We know it is possible for almost any sample to be drawn from almost any distribution; the only constraint being that $f_i$ at the location of each particle is non-zero, so the map between distributions and particles must be many-to-many---many distributions may be connected to one sample, and many samples may be sampled from one distribution. 

What I present now is a method of unbiasedly connecting distributions and samples.

Given that I have made a choice of $f$, the probability density of sampling a sequence of particles is found in equation \eqref{eq:NptProbDens}.

However, this equation does not reflect our ignorance with respect to $f$; I do this by introducing the joint probability that $f$ is chosen, and $\{\vw_i\}$ is obtained via taking $\tilde{N}$ Poisson samples from $f$,
\[
P_J = P[f] f^{(N)}(\{\vw_i\}) = P[f]\prod_{i = 1}^N f_i.
\]
Note that $\tilde{N}$ does not appear explicitly in $f^{(N)}$, because the number of samples taken is only probabilistically connected to the number of particles obtained via sampling. Thus $\tilde{N}$ has to be prescribed via a bootstrap method, which I now describe.

The assumption that the sample $\{\vw_i\}$ is randomly sampled implies that $P[f]$, the probability that $f$ is chosen, cannot be conditioned on $N$, the number of particles, only on $\tilde{N}$.

A system that is realised from Poisson sampling $\tilde{N}$ times has an expected probability of a sample that can be obtained via applying the law of large numbers,
\[\begin{aligned}
\prod_{i=1}^{\tilde{N}\mu} f_i &= \exp\Bigg(\sum_{i=1}^{\tilde{N}\mu} \ln f_i\Bigg) \rightarrow \exp\Bigg(\tilde{N}\int \d^6\vw ~f \ln f\Bigg)\\& = \exp(-\tilde{N}S),
\end{aligned}\]
where $S$ is the Shannon entropy of the probability density $f$,
\[
S[f] = - \int \d^6\vw~ f \ln f.
\]
This expected probability density of a sample obtained from an $\tilde{N}$-fold Poisson sampling defines the density of a sample that is definitively not an outlier, what \cite{Shannon1948} called a typical sample of $f$, $\{\vw_i\}_T$, 
\[\label{eq:TypCons}
f^{(N)}\Big(\{\vw_{i,T}\}\Big) = \prod_i (f_{i,T}) = \exp(-\tilde{N}S[f])
\]
Typical samples are samples that are expected (and thus, representative) of distributions $f$---but there are many $\{\vw_{i,T}\}$ which are typical of a single $f$ and many $f$ which are typical of one choice of $\{\vw_{i,T}\}$. The typicality constraint (equation \eqref{eq:TypCons}) already encodes indifference towards choosing any one choice of $\{\vw_{i,T}\}$ for $f$ by making each typical sample equiprobable. I then encode indifference towards choosing any one choice of $f$ for $\{\vw_{i,T}\}$, and this is done by asserting,
\[\label{eq:unbiasedness}
P_J[f,\{\vw_i\}_T] = P[f]\exp(-\tilde{N}S) = {1 \over \mathcal{Z}}
\]
where $\mathcal{Z}$ is a normalisation constant selected such that
\[
\int \mathcal{D}f~P[f] = 1,
\]
which will be enforced in the next section through a constrained entropy maximisation method. Equation \eqref{eq:unbiasedness} describes an unbiased prior on the space of distributions, which is equivalent to defining $P_J[f,\{\vw_i\}_T] = \mathcal{Z}^{-1}$ as the probability of a typical state, that is also understood in the language of Boltzmann's statistical mechanics as one of the microstates that contribute to a real system. 

The joint probability, $P_J[f,\{\vw_i\}]$, is identified as a thermodynamic probability: a probability that is proportional to the number of real microstates that characterise a physical system, the proportionality constant being $\mathcal{Z}^{-1}$, the probability of a single typical state.

The probability that $f$ is chosen must therefore take the form,
\[\label{eq:P[f]}
P[f] = {1 \over \mathcal{Z}}\exp(\tilde{N}S[f])
\]
if we are to unbiasedly connect $\{\vw_i\}$ and $f$.

Finally, I posit the core assumption of this theory, which also underlays the transition between Liouville's equation and the Collisionless Boltzmann Equation: that the observed sample of $N$ particles is indeed a typical sample obtained from an $\tilde{N} = N$-fold sampling of $f$. This gives us,
\[\label{eq:P_J}
P_J[f,\{\vw_i\}] = {1 \over \mathcal{Z}}\exp(NS[f])\prod_{i=1}^N f_i.
\]
This assumption encodes how the model $f$ connects to the samples $\{\vw_i\}$: $f$ is a model of the one-particle number density, and it must be asserted by fiat, because one can just as easily choose $f$ to be a model of the $2.2$ or $\fracj12$-particle number density. 

A demonstration of how $P_J$ discriminates between the pairs of $f,\{\vw_i\}$ is provided in Appendix \ref{AppendixBoxes}.

\subsection{Entropy Maximisation Criterion}\label{sec:EMaxCr}

The imposition of equation \eqref{eq:unbiasedness} is equivalent to assuming that $P_J[f,\{\vw_i\}]$ is flat in some space---this is also equivalent to maximising an entropy. In this section, I nail down this joint entropy maximisation criterion.

We begin with $P_J$ obtained via an $\tilde{N} = N$-fold sampling of $f$,
\[
P_J = P[f] \prod_{i = 1}^{N\mu} f_i
\]
We then define the joint entropy,
\[\begin{aligned}
S_J &= \int \mathcal{D}f\int...\int \prod_{i=1}^{N\mu} \d^6\vw_i~ \Bigg(-P_J\ln P_J\Bigg)\\
&= \int \mathcal{D}f~ \Bigg(-P[f]\ln P[f] \mu^{N\mu} +P[f] N\mu\mu^{N\mu-1}S[f]\Bigg)\\
&= \int \mathcal{D}f~\Bigg(-P[f]\ln P[f] \mu^{N\mu} + P[f] N\mu^{N\mu} S[f]\Bigg)
\end{aligned}\]
We reclaim equation \eqref{eq:P[f]} if we choose to extremise the joint entropy with respect to $P[f]$ while fixing $f$,
\[
{\delta S_J \over \delta P[f]}\Bigg|_{f} = 0
\]
which produces,
\[
P[f] \propto \exp(NS[f]).
\] 

This joint entropy maximisation criterion can be extended to include constraints on the expected sample obtained from $f$. One constraint that is of interest is the energy constraint, that we interpret as a constraint on the energy of the expected sample---which is also a constraint on the energy of the typical sample. 
\[\begin{aligned}\label{eq:EConstraint}
&\int \mathcal{D}f~\int \d^{6N\mu}\vw~ P_J \Bigg[H^{(N\mu)}(\{\vw_i\})-U\Bigg]\\& =  \int \mathcal{D}f~P[f]\mu^{N\mu}(E[f] - U)\\&=0,
\end{aligned}\]
defining $H^{(N\mu)}$ as the Hamiltonian of the expected sample, and $U$ the internal energy of the system. 

I introduce the ``normalisation constraints'',
\[\label{eq:Norm1}
\int \mathcal{D}f~\int ... \int \prod_{i=1}^{N\mu} \d^6\vw_i~ P_J = \Omega_N
\]
and
\[\label{eq:Norm2}
\int \mathcal{D}f~\int ... \int \prod_{i=1}^{N\mu} \d^6\vw_i~P_J \mu = \Omega_{N+1}
\]
where the former is a normalisation constraint for $P[f]$ and the latter's use will become apparent after entropy maximisation is executed. The newly introduced variable $\Omega_N$ is the total amount of probability for all possible phase-space configurations of particles averaged over all possible $f$ under the constraints utilised; i.e. it is the thermodynamic probability of the macrostate. We have assured ourselves that this is true because $P_J$ is a thermodynamic probability normalised to a typical microstate.

Application of the joint entropy maximisation criterion under these constraints via the method of Lagrange multipliers produces,
\[\label{eq:P_Edefn}
P_E[f] = \exp{(-1 + N(S[f]-\beta E[f] - \beta_\mathcal{Z} - \beta_\mu \mu))}
\]
where $E[f]$ is the self consistent energy (equation \eqref{eq:SelfConsistentE}) and the subscript $P_E$ denotes that an energy constraint has been introduced. This entropy maximisation criterion allows us to put forth a distribution of models where we have fixed certain macroscopic quantities, such as $N,V,U$, respectively, the number of particles, spatial volume and internal energy. Since any characteristic of the model $f$ may be obtained by taking moments of $P_E[f]$, I identify $P_E[f]$ here as a macrostate in an $N,V,U$ ensemble.

The Lagrange multipliers $\beta$, $\beta_\mu$ and $\beta_\mathcal{Z}$ are chosen to respect the energy constraint, a constraint that will be chosen later in the next section, and $\exp(1-N\beta_\mathcal{Z}) = {1/\mathcal{Z}}$ respectively. The method of Lagrange multipliers is detailed in Appendix \ref{AppendixMultipliers} alongside the workings of the energy constraint. 

\section{Ensemble Averages}\label{sec:EnsembleAverages}

Now that we have a way to distribute distribution functions unbiasedly, we turn to the question of actually computing ensemble averages with respect to $P[f]$, so as to facilitate the computation of macroscopic parameters by taking moments of $P[f]$, the macrostate.

Heuristically, we would like to compute ensemble averages by integrating over the space of all typical distribution functions of the observed sample, so as to produce the average DF that is typical of the sample---the ``correct distribution function''. 

This can also be achieved by computing the ensemble average with respect to $P[f]$, a result owed to \cite{Shannon1948} who proved that the typical samples of a distribution, while occupying a vanishing fraction of total configuration space, grow to dominate all of probability space as the number of samples taken increases---that is, any sufficiently large sample drawn from a distribution $f$ is highly likely to be a typical sample of that distribution. The typical distribution functions thus dominate the joint probability $P_J$, and so an unbiased integral (that is what $P[f]$ produces) over all $f$ is equal to integration over all typical $f$s.

We introduce the notation
\[
\langle G \rangle_{E} = \int \mathcal{D}f~P_E[f] G[f]
\]
that describes an ensemble average of a functional $G[f]$ taken with respect to $P_E[f]$. 

Computing these averages necessitates the use of perturbative field theory. We begin by introducing the separation
\[
f = f_0 + \delta f
\]
with $f_0$ to be defined a posteriori. 

Substituting into $P_E[f]$ we find,
\[\begin{aligned}\label{eq:P_EExpansion}
P_E&[f_0 + \delta f] = {1 \over \mathcal{Z}}\exp\Bigg(N\Bigg(S[f_0] - \beta E[f_0] - \beta_\mu \mu[f_0] \\&+ \int \d^6\vw_a~{\delta (S - \beta E - \beta_\mu \mu) \over \delta f_a}\Bigg|_{f = f_0} \delta f_a\\
&+ {1\over 2}\int \d^6\vw_a\d^6\vw_b~{\delta^2 (S - \beta E - \beta_\mu \mu) \over \delta f_a \delta f_b}\Bigg|_{f = f_0} \delta f_a \delta f_b\\&+ ....\Bigg)
\end{aligned}\]
We set the argument of the integral in the second line of \eqref{eq:P_EExpansion} to zero, thus defining $f_0$,
\[\label{eq:POME}
{\delta (S - \beta E - \beta_\mu \mu) \over \delta f}\Bigg|_{f = f_0} = 0
\]
This is the traditional principle of maximum entropy as employed by Gibbs, and implies,
\[
f_0 = \exp(-\beta H_0-1-\beta_\mu)
\]
where $H_0$ is the self-consistent Hamiltonian of $f_0$, and $f_0$ is an isothermal Hamiltonian. We now define the Lagrange multiplier $\beta_\mu$ to normalise $f_0$, such that:
\[
f_0 = C\exp(-\beta H_0),
\]
and
\[\label{eq:NormBetaMu}
C = \Bigg(\int \d^6\vw~ \exp(-\beta H_0)\Bigg)^{-1}
\]

The zeroth term in the expansion presented in equation \eqref{eq:P_EExpansion} is a constant, and the first order term has been set to zero by a judicious choice of $f_0$. The second order term is thus the dominant term; and we can write
\[\begin{aligned}
P_E[f_0 + \delta f] &= {1 \over \mathcal{Z}}\exp\Bigg({N \over 2} \int \d^6\vw_a \d^6\vw_b ~ \\&\times{\delta^2 (S - \beta E - \beta_\mu \mu) \over \delta f_a \delta f_b}\Bigg|_{f = f_0} \delta f_a \delta f_b\\
&+ N\sum_{n = 2}^N {(-1)^n \over n(n+1)} \int \d^6\vw  {\delta f^{n+1} \over f_0^n}\Bigg)
\end{aligned}\]
absorbing the first two orders into $\mathcal{Z}$. The subsequent terms in the expansion after the second order term originate (in this theory, with the energy constraint) purely from the Taylor expansion of $-f \ln f$ about $f_0$.

We now approximate the integral with respect to $f$ via Laplace's method. Those who are familiar with Stirling's approximation will find this method of approximating what is effectively the exponent of a sum of logarithms familiar. 

The conditions for Laplace's method is for the integrand to be (to leading order) a Gaussian with a large number $N$ in the argument. That this is true is owed to the nature of our constraints: the typicality constraint chooses a narrow range of $f$ for a given $\{\vw_i\}$, and the energy constraint only further narrows this range.

The approximation provided by Laplace's method improves as $N$ increases---this method of calculation is well suited to this statistical mechanical theory in which we probe large $N$ dynamics. Then we can neglect the higher ordered terms and truncate the expansion of $P_E$ at the quadratic term.

We first concern ourselves with computing the naked two-point correlation function,
\[
\langle \delta f \delta f' \rangle_E = \int \mathcal{D}f~P[f_0 + \delta f] \delta f \delta f'.
\]
To leading order,
\[\begin{aligned}
\langle \delta f \delta f' \rangle_E = \int \mathcal{D}f~&{1 \over \mathcal{Z}} \exp\Bigg({N\over 2} \int \d^6\vw_a \d^6\vw_b~\\&{\delta ^2(S - \beta E - \beta_\mu \mu) \over \delta f_a \delta f_b}\Bigg|_{f = f_0}\delta f_a \delta f_b\Bigg)\delta f \delta f'
\end{aligned}\]
We solve for the related quantity;
\[\begin{aligned}
&-\Bigg\langle \int \d^6\vw_a~\delta f N{\delta^2 (S-\beta E - \beta_\mu\mu) \over \delta f' \delta f_a}\Bigg|_{f = f_0} \delta f_a\Bigg\rangle_E\\
&\approx -\int \mathcal{D}f~ \delta f {\delta P_E[f] \over \delta f'}\\
&= \delta^6(\vw - \vw')
\end{aligned}\]
where the first equality holds approximately, neglecting the third and higher order terms in the expansion, the second equality is obtained via functional integration by parts in conjunction with the definition for functional derivatives (equation \eqref{eq:FunctDerivDef}) and the normalisation of $P[f]$. This result implies that
\[\label{2-pt-correlation}
\langle \delta f \delta f'\rangle_E = -{1 \over N} \Bigg({\delta ^2 ( S - \beta E - \beta_\mu \mu) \over \delta f \delta f'}\Bigg)^{-1}\Bigg|_{f = f_0} + ...
\]
where the next correction is $\propto 1/N^2$ and the inverse of a function $K(\vw,\vw')$ is defined by,
\[\label{eq:InverseDefinition}
\int \d^6\vw_a~ K(\vw,\vw_a) K^{-1}(\vw_a,\vw') = \delta^6(\vw - \vw').
\]

I will now illustrate how we compute this inverse.

Taylor expanding, we find:
\[\label{eq:S-bE}
-{\delta^2 (S - \beta E - \beta_\mu \mu) \over \delta f \delta f' } \Bigg|_{f = f_0} = {1 \over f_0}\delta^6(\vw - \vw') - {G Mm \beta \over |\vx - \vx'|}
\]
Let us now define the ``matrices'',
\[
A = {1 \over f_0}\delta^6(\vw - \vw')
\]
and
\[
B = -{G Mm\beta \over |\vx - \vx'|}.
\]
We use equation \eqref{eq:InverseDefinition} to find,
\[
A^{-1} = f_0 \delta^6(\vw - \vw')
\]
and can now apply the expansion (really, just the Taylor expansion of $1/(1+x))$,
\[
(A + B)^{-1} = A^{-1}(I - BA^{-1} + (BA^{-1})^2 - (BA^{-1})^3 + ...)
\]
which is only applicable if the spectral norm (magnitude) of $||B|| << ||A||$. We recall that ``matrix multiplication'' really means integration over shared variables to find that
\[\begin{aligned}\label{eq:A+Binv}
(A+B)^{-1}(\vw,\vw') &= f_0 \delta^6(\vw - \vw') + f_0 f_0'\Bigg({GMm\beta \over |\vx - \vx'|}\\& + \int \d^6\vw_a~ {GMm\beta \over |\vx - \vx_a|}f_{0a}{GMm\beta \over |\vx_a - \vx'|} + ...\Bigg)\\
&= f_0 \delta^6(\vw - \vw') + f_0 f_0' X(\vx,\vx').
\end{aligned}\]
The spatial correlation function $X(\vx,\vx')$ is,
\[
X(\vx,\vx') = {GMm\beta \over |\vx - \vx'|} + \int \d^6\vw_a~{GMm\beta \over |\vx - \vx_a|}f_{0a} {GMm\beta \over |\vx_a- \vx'|} + ...
\]
We apply the Laplacian to obtain a PDE for $X$,
\[\begin{aligned}
\vnabla^2 X(\vx , \vx') &= -4\pi G Mm\beta \delta^3(\vx - \vx') - 4\pi GMm\beta\rho_0\Bigg({GMm\beta \over |\vx - \vx'|}\\
&+  \int \d^6\vw_a~{GMm\beta \over |\vx - \vx_a|} f_{0a} {GMm\beta \over |\vx_a - \vx'|} + ...\Bigg)\\
&= -4\pi GMm\beta \delta^3(\vx - \vx') - 4\pi GMm\beta \rho_0 X(\vx,\vx')\\
\end{aligned}\]
which is valid even beyond the boundaries of the matrix expansion above,
\[\label{eq:SpatialCorrDefn}
\Bigg(-{1\over 4\pi GMm\beta} \vnabla^2 - \rho_0\Bigg)X(\vx,\vx') = \delta^3(\vx - \vx').
\] The physical interpretation of this result can be obtained via a re-organisation of terms,\[
{1 \over 4\pi G} \nabla^2 (-X(\vx,\vx')/M\beta) = {m\rho_0}X(\vx,\vx') + m\delta^3(\vx - \vx')
\] is Poisson's equation describing the potential induced by a particle of mass $m$ and the polarisation cloud it induces on the background medium. Here, $\rho_0 = \int \d^3\vv ~f_0$ is the spatial distribution function. 

Substituting equation \eqref{eq:A+Binv} into equation \eqref{2-pt-correlation} gives us the two point correlation function, which is normalised to $1$;
\[\label{eq:2ptCorrelationE}
\langle \delta f \delta f'\rangle_E = {1\over N}\Bigg(f_0\delta^6(\vw - \vw') + f_0 f_0' X(\vx,\vx')\Bigg).
\]
This two point correlation function equation mirrors that found by \cite{Bose2023}, who computed their ``two-particle correlation function'' studying the steady-state solution of the BBGKY hierarchy truncated at the third order. My calculation, however, highlights that the polarisation potential and density are self-similar.

Let us now gain some intuition into this result by applying it to some well-known examples.

\subsection{The Maxwellian}

Perhaps the best known example of an isothermal system is that that is characterised by the Maxwellian distribution\[\label{eq:Maxwellian}
f_0 = C\exp({-\fracj12 m\vv^2}).
\]
It is spatially homogeneous---seemingly ill-suited to a gravitating system in which attractive forces exacerbate inhomogeneities. However, Jeans showed that a ``Maxwellian trapped in a sphere'' could be stable if the sphere was sufficiently small compared to a physical quantity known as the Jeans' length and was immersed in an infinite, static density medium with the same density as $f_0$.

We now compute the two-point correlations for a self-gravitating Maxwellian trapped in such a sphere of radius $r_m$, beyond which is an infinite static medium. 

Due to the normalisation $\int \d^6\vw~ f_0 = 1$, we obtain $\rho_0 = 1/V_\vx$ where $V_\vx = \fracj43\pi r_m^3$. We also recall that $M = N*m$.

Substituting into equation \eqref{eq:SpatialCorrDefn}, the spatial correlation function obeys \[\label{eq:SpatialCorrDefnMax}\Bigg(- \vnabla^2 - {4\pi GMm\beta \over V_\vx}\Bigg)X(\vx,\vx') = 4\pi GMm\beta\delta^3(\vx - \vx')\]
Let us now define the inverse length scale, the Jeans wavenumber $k_J$,
\[
k_J^2 = {4\pi GMm \beta \over V_\vx}
\]
and the Fourier transform of $X(\vx,\vx')$, $\mathcal{F}[X]$, noting that the symmetry of this PDE allows us to write $X = X(\vx-\vx')$.
\[
X(\vx - \vx') = {1 \over (2\pi)^3}\int \d^3\vk ~\exp(\i \vk \cdot (\vx - \vx')) \mathcal{F}[X](\vk)
\]
Then, the Fourier transform of equation \eqref{eq:SpatialCorrDefnMax} is,
\[
(\vk^2 - k_J^2)\mathcal{F}[X](\vk) = 4\pi GMm\beta
\]
and the inverse transform produces
\[
X(\vx - \vx') = {1 \over (2\pi)^3}\int \d^3\vk~ \exp(\i\vk\cdot(\vx - \vx')) {4 \pi GMm\beta \over \vk^2 - k_J^2}.
\]
Solving this integral is a textbook exercise;
\[
X(|\vx - \vx'|) = {GMm\beta}{\cos(k_J |\vx - \vx'|) \over |\vx - \vx'|}.
\]
Thus we find that the correlations seeded by the gravitational interactions are long-ranged and promote the ``clustering'' of particles. 

We can find the correlations for an equivalent electrostatic system where particles are of charge $q$, the total charge of the system is $Q$, and the permissivity of free space is $\epsilon_0$ by taking the map,
\[
GMm \rightarrow -{q Q \over 4\pi \epsilon_0}
\]
which sends $k_J \rightarrow \i k_D$, the Debye wavenumber and produces a spatial correlation function $X_e$,
\[
X_e(|\vx - \vx'|) = -{qQ \beta\over4\pi\epsilon_0}{\exp(-k_D |\vx-\vx'|) \over |\vx-\vx'|}.
\]
This is a statement of \cite{Debye1923} shielding; the short-ranged correlations (as evident from the exponential damping) ensures that distant groups of charged particles are uncorrelated with each other, and the correlations that do develop are negative---that is, the particles repel.

\subsection{Background Correlation Functions}

A surefire way to understand the correlation range of a system is to compute its ``total correlation''. To the lowest order, we can do this by computing the average two-point correlation between a particle and its $ \approx N$ peers. 

I introduce the dimensionless background correlation function (BCF), $\xi$:
\[\label{eq:BGCorrelation}
\xi_E(\vw) = {N \over f_0}\int \d^6\vw_a~{\langle \delta f(\vw) \delta f_a\rangle_E}
\]
that represents the naked two-point correlations between a particle and every other particle in the distribution---note the phase-space integral spans available phase-space, which is $r \leq r_m$ in this example. Substituting equation \eqref{eq:2ptCorrelationE} into equation \eqref{eq:BGCorrelation},
\[
\xi_E(\vx) = \Bigg(1 + \int \d^6\vw_a f_{0a}X(\vx,\vx_a)\Bigg) 
\]
We integrate equation \eqref{eq:SpatialCorrDefn} with respect to $f_0' \d^6\vw'$ to find,
\[\begin{aligned}
\int \d^3\vx_a~ \Bigg(-{\vnabla^2\over 4\pi G Mm\beta}X(\vx,\vx_a) \rho_{0a} - \rho_{0} X(\vx,\vx_a) \rho_{0a} \Bigg) = \rho_0.
\end{aligned}\]
After some algebraic manipulation, we find
\[
\Bigg(-{\vnabla^2 \over 4\pi G Mm\beta}-\rho_0\Bigg)\xi_E(\vx) = 0
\]
With two boundary conditions: the BCF is finite as $\vx \rightarrow 0$ and takes on a well-defined value at $\vx \rightarrow 0$, that can be computed by direct evaluation of $\xi_E(0)$. We substitute the Maxwellian DF, 
\[\begin{aligned}
\xi_E(0) &= 1 + \int \d^6\vw~ f_{0} X(0,\vx)\\
&= k_J r_m \sin(k_J r_m) + \cos(k_J r_m).
\end{aligned}\]
or,
\[
\xi_E(\vx) = \Bigg(k_J r_m \sin(k_J r_m) + \cos(k_J r_m)\Bigg) \mathrm{sinc}(k_J r)
\]
Note that the electrostatic BCF can be computed via the same procedure, 
\[
\xi_{E,e}(\vx) = \Bigg((1+ k_D r_m) \exp(-k_D r_m)\Bigg){\mathrm{sinh}(k_D r)\over k_D r}
\]
The upper bound of $\mathrm{sinc}(k_J r)$ is $1$, while the upper bound of $\mathrm{sinh}(k_D r) / k_D r$ is $\mathrm{sinh}(k_D r_m)/k_D r_m$ since $r = r_m$ is the maximum radius at which particles can be located. It is evident that the maximum of $\xi_E$ is unbounded, growing as $r_m \rightarrow \infty$, while the maximum value of $\xi_{E,e}$ approaches $1$ taking the same limit. 

This unbounded growth in the ``total correlation'' for gravitating systems will be explored in the next paper in this series.

\section{Discussion}

In this section, I describe the results and findings of this paper.

\subsection{The Probability Volume of the Macrostate, $\Omega_N$}

In section \ref{sec:EMaxCr} I introduced a pair of normalisation constraints into the theory, equations \eqref{eq:Norm1} and \eqref{eq:Norm2}. 

Respectively, they represent the normalisation of the expected distribution function of particles obtained from an $N$-fold sampling, and the normalisation of the expected distribution function of particles plus one additional particle. 

These normalisation constraints, if set to $1$, would ensure that adding particles to the system (while generating new correlations) would not change the total amount of probability occupied by all possible configurations of the system---this is equivalent to fixing the total probability of the macrostate. 

I instead chose to adhere to the statistical mechanical approach: fixing the probability of a typical microstate, and allowing the thermodynamic probability of the macrostate to fluctuate in response to fluctuations in the number of typical microstates.

This ensures that correlations beyond the maximum entropy state are represented by variations in $\Omega_N$. In a later publication, I will show that  variations in $\Omega_N$ can be connected to basic thermodynamical quantities.

\subsection{Normalisation}

Perhaps the most unorthodox choice in the writing of this paper is the choice to depart from normalised DF, $f$, in favour of its interpretation as a number density, albeit treated probabilistically via the Poisson sampling method. The choice to depart from a normalised probability distribution function is not without precedence, however, as it was first chosen by \cite{Boltzmann1877}, who also implicitly abandoned a normalised DF in favour of fixing the probability of a real microstate.

The nature of Poisson sampling is such that the total amount of probability (that is, the probability of successfully finding particles, added to the probability of not finding particles, summed over each point in phase-space) for a single Poisson sampling is $\int \d^6\vw~f + (1 - \d^6\vw~f) = \int 1$ which is formally undefined: this is the total density associated with finding a particle at every point in the phase-space continuum. If we are to normalise $f$, we must normalise it with respect to the total amount of probability available to sample, which is $\int 1$. It is evident that it doesn't matter whether $\mu = \int \d^6\vw~f = 1$ or $2$ then. 

This result also contextualises the probability of the macrostate, $\Omega_N$. Macrostates of different $N,V,U$ and of different constraints all occupy the same uncountably infinite probability space, which is able to account for the sampling of every particle at all points in phase-space. 

\subsection{The Ergodic Hypothesis and Typicality}

The ergodic hypothesis states that a system explores all the phase-space available to it evenly. Instead of averaging over the accessible phase-space configurations of the sample, as the ergodic hypothesis implies, I average over the space of distributions.

As stated in the beginning of section \ref{sec:EnsembleAverages}, this average over the space of distributions is equivalent to averaging over all typical distribution functions of a given sample. 

Thus, I replace the ergodic hypothesis with Shannon's typicality criterion, fitting the sample with typical $f$. The unbiasedness assumption (equation \eqref{eq:unbiasedness}) states that pairs of typical $f,\{\vw_i\}$ are equiprobable---this is the successor to the ergodic hypothesis, and typical samples replace the surface of constant energy.

Boltzmann hypothesises that in the limit of large $N$, the entropy maximising distribution function $f_0$ (i.e. his Boltzmann distribution) contributes dominantly to the production of real microstates, which are all equally likely. In my model, all pairs of $f,\{\vw_i\}$ contribute via $P[f]$ to the production of real microstates. 

Showing that the entropy maximising distribution function $f_0$ does not always dominate the production of real microstates a priori of dynamics even in the limit of large $N$ is the subject of the next paper, but a heuristic argument is as follows. Equation \eqref{eq:P_J} shows that any calculation of the ensemble averaged probability of a sample $\langle \prod_i f_i \rangle$ requires taking the $N$-th uncentred moment of $P[f]$. If $P[f]$ is fully Gaussian, then we find $\langle \prod_i f_i \rangle = \prod_i f_{0i}$. $P[f]$ is only Gaussian to the $1/N$-th order, however, so we should expect correction terms to be of order unity, $N * 1/N = 1$, precisely as the calculation of the background correlation functions imply.

\subsection{Ensemble Averages and Typicality}

The ensemble averages presented in this paper integrate over the space of models, $f$, and seem to differ from the usual meaning of an ensemble average. I will now argue that they are the generalisation of the usual interpretation of an ensemble average.

As highlighted in the introduction, an ensemble average in Gibbs' thermodynamics is positioned to replace a time-average; instead of averaging over the deterministic but chaotic sequence of states produced by time-evolution, Gibbs suggests that we average over a distribution of dynamically accessible states.
\[
\bar{f}(\vw) = \int_0^{T} \d t~f(\vw,t)
\]

Another example of when ensemble averages are used is to represent an angle average---that is, an average over the angle variable $\vtheta$ of the action-angle coordinates $\vw = (\vtheta,\vJ)$. As $\vtheta$ represents the position of a particle along the trajectory parametrised by $\vJ$, angle averaging is akin to taking an average of $f(\vw,t)$ over all possible positions of the particle along its trajectory, weighted according to the time spent by the particle at each segment in its trajectory; reducing \[
\bar{f}(\vJ,t) = \int \d^3\vtheta~ f(\vtheta,\vJ,t).
\]For a spatially homogeneous plasma, this angle-average is just a spatial average,\[
\bar{f}(\vv,t) = \int \d^3\vx~f(\vx,\vv,t).
\]
These conventional ensemble averages implicitly average over different distribution functions by averaging the DF over coordinates not thought to be important to the dynamics of interest---angle averaging averages over angle coordinates and is often used to investigate secular dynamics (i.e., dynamics on timescales that far exceed the orbital timescale), where the forces at play are thought to `apply evenly at all angles'. Time-averaging averages over microscopic physics that is `smoothed over' when macroscopic features are measured. Not only does my ensemble average describe this averaging explicitly without requiring judgement on which coordinates are non-essential to the dynamics of the system; it also provides the fundamental statistics that justifies these averaging procedures, that is the averaging over instances of the distribution function thought to be statistically equivalent, be it by fiat or by experimental evidence.

By checking if $P_E[f]$ contains no explicit time-dependence, that is ${\p \over \p t}P_E[f] = 0$, we can reclaim time-stationary statistics: this also opens the way to introducing explicitly time-varying statistics.

\subsection{Lagrange Multipliers}

The thermodynamic $\beta$ is the Lagrange multiplier used to enforce an energy constraint, and is used in the definition of the Jeans and Debye wavenumbers. 

However, $\beta$ is designed to enforce the constraint,\[
\langle \mu^{N\mu} N(E[f]) \rangle = U\langle \mu^{N\mu} \rangle.
\] $\beta$ is constrained not just on the kinetic energy of the system, but on its potential energy. We can rework this result to a more suggestive form:
\[
-{\p \over \p \beta}\ln(\mathcal{Z}\langle \mu^N \rangle/N!) = U,
\]
which is in the form of a Boltzmann entropy maximisation criterion,
\[
-{\p \over \p \beta}\ln W = U,
\]
that is the definition of $\beta$. Showing that this definition of $\beta$ is a generalisation of the standard thermodynamic beta will be left to the third paper in this series.

This theory produces both the Boltzmann entropy maximisation criterion that governs thermodynamics, and the $S[f]$ entropy maximising criterion that describes the entropy-maximising distribution function as mathematical corollaries of the typicality assumption.

\subsection{Covariance with Dynamics}

The CBE, while not actively invoked in the writing of this paper that is wholly concerned with attributing probabilities to samples and distribution, is a constant consideration.

The use of canonical phase-space coordinates $\vw = (\vx,\vv)$ is essential: for a probability measure on the space of distributions and samples to be useful, it must assign the same probability to a system at any point in time of its evolution---thus being able to assign probabilities not just to a sample of particles, but also the entire trajectory these particles trace in time \citep{LauBinney_probDF}. That is to say that,
\[
{\d \over \d t}\prod_i f(\vw_i,t) = 0
\]
which is true given that $f$ evolves under the CBE. 

Perhaps a more interesting question is the time-evolution of quantities like $\langle {\p/\p t} f_i \rangle$. The dynamics of ensemble averaged macroscopic variables will come into focus when we do compute them in the next paper in this series. 

\section{Conclusions}

In this paper, I present a means of representatively modelling a given sample, incorporating certain qualities that are expected of said sample. These models naturally form an ensemble of ``suitable fits'' to the sample.

I posit that computing ensemble averages should be done over the space of distributions, and that conventional means of ensemble averaging average over the space of distributions implicitly. 

Samples and distributions are connected in an unbiased manner via Shannon's entropy, which has been extended to describe the Poisson sampling of a density $f$, and a means of constraining the probability distribution of $f$ based on the qualities of the expected sample is furnished. 

Two point correlations are computed from this theory, and familiar results are obtained, alongside a rederivation of Debye shielding. Finally, the background correlation function is defined, that describes lowest order fluctuations of order unity that arise from having two-point correlations develop between a particle and its $\approx N$ counterparts.

The next paper, titled `Correlation Functions', will furnish the computation of higher order correlation functions.

The second to next paper, titled `Statistical Mechanics', will describe how we can derive statistical mechanics from this theory. 

\section*{Acknowledgements}

I acknowledge Shanghai Jiao Tong University's Siyuan Postdoctoral Talent Programme and University College London's Graduate Research Scholarship for funding the research behind this paper. 

I also thank James Binney and Ralph Schoenrich, for many invaluable comments across the writing of this paper. I thank Alexander Schekochihin and Robert Ewart for beating the crap out of my half-baked theories. I thank Minghao Li for giving me the funniest idea over a pint---apply the mathematical formalism of QFT to astrophysical statistical mechanics---and I thank Thormund Tay for helpful discussions regarding the unbiased connection between $f,\{\vw_i\}$. I thank Kinwah Wu for his advice and thoughts regarding the lack of normalisation and the alternative to the principle of maximum entropy. Last but not least, I thank Douglas Heggie and Daniel Verscharen for their comments on my thesis, part of which entered this paper.

\section*{Data Availability}

 No data was generated in the writing of this paper.


\input{output_bbl.bbl}
\bibliographystyle{mnras}

\begin{thebibliography}{}
\makeatletter
\relax
\def\mn@urlcharsother{\let\do\@makeother \do\$\do\&\do\#\do\^\do\_\do\%\do\~}
\def\mn@doi{\begingroup\mn@urlcharsother \@ifnextchar [ {\mn@doi@} {\mn@doi@[]}}
\def\mn@doi@[#1]#2{\def\@tempa{#1}\ifx\@tempa\@empty \href {http://dx.doi.org/#2} {doi:#2}\else \href {http://dx.doi.org/#2} {#1}\fi \endgroup}
\def\mn@eprint#1#2{\mn@eprint@#1:#2::\@nil}
\def\mn@eprint@arXiv#1{\href {http://arxiv.org/abs/#1} {{\tt arXiv:#1}}}
\def\mn@eprint@dblp#1{\href {http://dblp.uni-trier.de/rec/bibtex/#1.xml} {dblp:#1}}
\def\mn@eprint@#1:#2:#3:#4\@nil{\def\@tempa {#1}\def\@tempb {#2}\def\@tempc {#3}\ifx \@tempc \@empty \let \@tempc \@tempb \let \@tempb \@tempa \fi \ifx \@tempb \@empty \def\@tempb {arXiv}\fi \@ifundefined {mn@eprint@\@tempb}{\@tempb:\@tempc}{\expandafter \expandafter \csname mn@eprint@\@tempb\endcsname \expandafter{\@tempc}}}

\bibitem[\protect\citeauthoryear{{Bernoulli}}{{Bernoulli}}{1738}]{Bernoulli1738}
{Bernoulli} D.,  1738, Hydrodynamica, sive de viribus et motibus fluidorum commentarii.
sumptibus Johannis Reinholdi Dulseckeri : Typis Joh. Deckeri, typographi Basiliensis

\bibitem[\protect\citeauthoryear{{Binney} \& {Tremaine}}{{Binney} \& {Tremaine}}{2008}]{GDII}
{Binney} J.,  {Tremaine} S.,  2008, {Galactic Dynamics: Second Edition}.
Princeton University Press

\bibitem[\protect\citeauthoryear{{Bogoliubov}}{{Bogoliubov}}{1946}]{Bogoliubov1946}
{Bogoliubov} N.~N.,  1946, Journal of Physics USSR.

\bibitem[\protect\citeauthoryear{{Boltzmann}}{{Boltzmann}}{1877}]{Boltzmann1877}
{Boltzmann} L.,  1877, Sitzungberichte der Kaiserlichen Akademie der Wissenschaften. Mathematisch-Naturwissen Classe

\bibitem[\protect\citeauthoryear{Born \& Green}{Born \& Green}{1946}]{BornGreen1946}
Born M.,  Green H.~S.,  1946, \mn@doi [Proceedings of the Royal Society of London. Series A. Mathematical and Physical Sciences] {10.1098/rspa.1946.0093}, 188, 10

\bibitem[\protect\citeauthoryear{Bose}{Bose}{2023}]{Bose2023}
Bose A.,  2023, The European Physical Journal B, 96, 41

\bibitem[\protect\citeauthoryear{Clausius}{Clausius}{1850}]{Clausius1850}
Clausius R.,  1850, Annalen der Physik, 155, 368

\bibitem[\protect\citeauthoryear{Debye \& Hückel}{Debye \& Hückel}{1923}]{Debye1923}
Debye P.,  Hückel E.,  1923, Physikalische Zeitschrift, 24, 185

\bibitem[\protect\citeauthoryear{{Gibbs}}{{Gibbs}}{1902}]{Gibbs}
{Gibbs} J.~W.,  1902, {Elementary Principles in Statistical Mechanics}.
Yale University Press

\bibitem[\protect\citeauthoryear{{Kirkwood}}{{Kirkwood}}{1946}]{Kirkwood1946}
{Kirkwood} J.~G.,  1946, The Journal of Chemical Physics

\bibitem[\protect\citeauthoryear{{Lau}}{{Lau}}{2024}]{LauJunYan2024}
{Lau} J.~Y.,  2024, PhD thesis, University College London

\bibitem[\protect\citeauthoryear{Lau \& Binney}{Lau \& Binney}{2021}]{LauBinney_probDF}
Lau J.~Y.,  Binney J.,  2021, \mn@doi [Monthly Notices of the Royal Astronomical Society] {10.1093/mnras/stab2047}, 506, 4007

\bibitem[\protect\citeauthoryear{{Liouville}}{{Liouville}}{1838}]{Liouville}
{Liouville} J.,  1838, Journal de mathématiques pures et appliquées

\bibitem[\protect\citeauthoryear{Metropolis \& Ulam}{Metropolis \& Ulam}{1949}]{MetropolisUlam1949}
Metropolis N.,  Ulam S.,  1949, \mn@doi [Journal of the American Statistical Association] {10.1080/01621459.1949.10483310}, 44, 335

\bibitem[\protect\citeauthoryear{Shannon}{Shannon}{1948}]{Shannon1948}
Shannon C.~E.,  1948, \mn@doi [The Bell System Technical Journal] {10.1002/j.1538-7305.1948.tb01338.x}, 27, 379

\bibitem[\protect\citeauthoryear{{Yvon}}{{Yvon}}{1935}]{Yvon1935}
{Yvon} J.,  1935, Actual. Sci. and Indust. (Paris, Hermann)

\makeatother
\end{thebibliography}




\appendix

\section{Box Distributions and Smoothening}\label{AppendixBoxes}

Intuition with respect to equation \eqref{eq:P_J} can be obtained by considering this experimental scenario: we have obtained a scattering of particles in phase-space, $\{\vw_i\}$, and we would like to extract information by binning the particles using a $6D$-histogram, comprised of regular phase-space hypercubes of volume $\Delta$, counting the number of particles, and then dividing by $N$. 

The candidate distribution function obtained (as a function of cube-size $\Delta$ and the number of particles within the cube, $n_c$) from this process is
\[
f(c,\Delta) = {n_c \over N \Delta}.
\]
Substituting $f(c,\Delta)$ into $P_J$ we find \[\begin{aligned}&P_J[f(c,\Delta),\{\vw_i\}] = {1 \over \mathcal{Z}}\exp(NS[f(c,\Delta)])\prod_c f(c,\Delta)^{n_c}\\
&={1 \over \mathcal{Z}}\exp\Bigg(NS[f(c,\Delta)] + N\Delta\sum_c f(c,\Delta) \ln(f(c,\Delta))\Bigg)\\ &= {1 \over \mathcal{Z}}\end{aligned}\]
where between the second and final equalities, I have used the definition of the Shannon entropy. 

This substitution reveals that histogram distribution functions, irrespective of the choice of $\Delta$, are typical of the sample $\{\vw_i\}$. This is true when $\Delta \rightarrow 0$ or when $\Delta \rightarrow \infty$, which are the Klimontovich distribution function and the maximally ignorant prior respectively. The fact that these distribution functions are equiprobable draws us in to the meaning of the typicality condition: A typical sample of $f$ is by definition a non-outlierly sample of $f$---and it is impossible to sample an outlier from either the Klimontovich DF, which is a sum of Dirac delta functions, or the maximally ignorant DF, which has a constant probability. 

In conclusion, $P_J$ discriminates only between distribution functions for which the sample can be said to be a good fit and distribution functions for which the sample cannot be said to be a good fit, without introducing additional information.

\section{Lagrange Multipliers on the Space of Distributions}\label{AppendixMultipliers}

The standard theory of Lagrange multipliers is as follows: given that we are trying to find the extremum point of the function $f(x)$ with respect to the constraints $g_i(x) = 0$,

We can extremise the Lagrangian,
\[
\mathcal{L} = f(x) + \sum_i \lambda_i g_i
\]
where the solutions to the equations
\[
{\p \mathcal{L} \over \p x} = 0;~{\p \mathcal{L}\over \p \lambda_i} = 0
\]
define the values of $x$ and $\lambda_i$ that maximise $f(x)$ under the aforementioned constraints.

In section \ref{sec:EMaxCr}, the energy constraint (equation \eqref{eq:EConstraint}) is
\[\begin{aligned}
\int \mathcal{D}f \int ... \int \prod_{i=1}^{N\mu} &\d^6\vw_i~ P_J \Bigg( H^{(N\mu)}(\{\vw_i\}) - U \Bigg) \\&= \langle \mu^{N\mu}(NE[f]-U)\rangle = 0.
\end{aligned}\]
and the normalisation constraints (equations \eqref{eq:Norm1} and \eqref{eq:Norm2}) are\[ \begin{aligned}
 \int \mathcal{D}f\int ... \int \prod_{i=1}^{N\mu} \d^6\vw_i~ P_J &= \langle \mu^{N\mu} \rangle = \Omega_N\\
 \int \mathcal{D}f\int ... \int \prod_{i=1}^{N\mu} \d^6\vw_i~ P_J \mu &= \langle \mu^{N\mu+1} \rangle = \Omega_{N+1} .
\end{aligned}
\]
The Lagrangian to be maximised $\mathcal{L} = \mathcal{L}[P,f]$ is\[
\mathcal{L} = S_J - \beta N\langle \mu^{N\mu} E[f]\rangle - \beta_\mathcal{Z} \langle \mu^{N\mu} \rangle - \beta_{\mu} \langle \mu^{N\mu + 1}\rangle
\]
so
\[\begin{aligned}
{\delta \mathcal{L} \over \delta P}\Bigg|_{f} &= \Bigg(-\ln P - 1 + N S[f] - \beta N (E[f]) -\beta_\mathcal{Z}  - \beta_\mu \mu\Bigg)\mu^{N\mu}\\
&= 0
\end{aligned}\]
which works out to
\[
P[f] = \exp(-1 + N(S-\beta E) - \beta_\mathcal{Z} - \beta \mathcal{\mu})
\]
A simple reparametrisation then reproduces equation \eqref{eq:P_Edefn}.


\bsp	
\label{lastpage}
\end{document}